\documentclass[traditabstract]{aa}
\usepackage{graphicx}
\usepackage{txfonts}

\newcommand{\secp}{\mbox{\rlap{.}$''$}} 
 
\newcommand{\tablenotea}[1]{\parbox{  8.9cm}{\indent \footnotesize{#1}}}

\newcommand{\nature}{Nature}

\newcommand{\cpl}{Chem. Phys. Lett.}
\newcommand{\cp}{Chem. Phys.}
\newcommand{\ieee}{IEEE Trans. Antennas Propag.}

\newcommand{\chemrev}{Chem. Rev.}

\newcommand{\jpc}{J. Phys. Chem.}

\newcommand{\psp}{Planet. Space Sci.}

\begin{document}

\title{Probing non polar interstellar molecules through their protonated form: Detection of protonated cyanogen (NCCNH$^+$)\thanks{Based on observations carried out with the IRAM 30m Telescope and the Yebes 40m Telescope. IRAM is supported by INSU/CNRS (France), MPG (Germany) and IGN (Spain). The 40m radiotelescope at Yebes Observatory is operated by the Spanish National Geographic Institute (IGN, Ministerio de Fomento).}}

\titlerunning{Interstellar protonated cyanogen}
\authorrunning{Ag\'undez et al.}

\author{
M.~Ag\'undez\inst{1}, J.~Cernicharo\inst{1}, P.~de~Vicente\inst{2}, N.~Marcelino\inst{3}, E.~Roueff\inst{4}, A.~Fuente\inst{5}, M.~Gerin\inst{6}, M.~Gu\' elin\inst{7}, C.~Albo\inst{2}, A.~Barcia\inst{2}, L.~Barbas\inst{2}, R.~Bola\~no\inst{2}, F.~Colomer\inst{2}, M.~C.~Diez\inst{2}, J.~D.~Gallego\inst{2}, J.~G\'omez-Gonz\'alez\inst{2}, I.~L\'opez-Fern\'andez\inst{2}, J.~A.~L\'opez-Fern\'andez\inst{2}, J.~A.~L\'opez-P\'erez\inst{2}, I.~Malo\inst{2}, J.~M.~Serna\inst{2} and F.~Tercero\inst{2}
}

\institute{
Instituto de Ciencia de Materiales de Madrid, CSIC, C/ Sor Juana In\'es de la Cruz 3, 28049 Cantoblanco, Spain \and
Centro Nacional de Tecnolog\'ias Radioastron\'omicas y Aplicaciones Geoespaciales(CNTRAG), Observatorio de Yebes (IGN), Spain \and
INAF, Istituto di Radioastronomia, via P. Gobetti 101, 40129 Bologna, Italy \and
LERMA, Observatoire de Paris, PSL Research University, CNRS, UMR8112, Place Janssen, 92190 Meudon Cedex, France \and
Observatorio Astron\'omico Nacional (OAN), Calle Alfonso XII, No 3, 28014 Madrid, Spain \and
LERMA, Observatoire de Paris, \'Ecole Normale Sup\'erieure, PSL Research University, CNRS, UMR8112, F-75014, Paris, France \and
Institut de Radioastronomie Millim\'etrique, 300 rue de la Piscine, 38406 St. Martin d'H\'eres, France
}

\date{Received; accepted}


\abstract
{Cyanogen (NCCN) is the simplest member of the series of dicyanopolyynes. It has been hypothesized that this family of molecules can be important constituents of interstellar and circumstellar media, although the lack of a permanent electric dipole moment prevents its detection through radioastronomical techniques. Here we present the first solid evidence of the presence of cyanogen in interstellar clouds through the detection of its protonated form toward the cold dark clouds TMC-1 and L483. Protonated cyanogen (NCCNH$^+$) has been identified  through the $J=5-4$ and $J=10-9$ rotational transitions using the 40m radiotelescope of Yebes and the IRAM 30m telescope. We derive beam averaged column densities for NCCNH$^+$ of $(8.6 \pm 4.4) \times 10^{10}$~cm$^{-2}$ in TMC-1 and $(3.9 \pm 1.8) \times 10^{10}$~cm$^{-2}$ in L483, which translate to fairly low fractional abundances relative to H$_2$, in the range (1-$10)\times10^{-12}$. The chemistry of protonated molecules in dark clouds is discussed, and it is found that, in general terms, the abundance ratio between the protonated and non protonated forms of a molecule increases with increasing proton affinity. Our chemical model predicts an abundance ratio NCCNH$^+$/NCCN of $\sim10^{-4}$, which implies that the abundance of cyanogen in dark clouds could be as high as (1-$10)\times10^{-8}$ relative to H$_2$, i.e., comparable to that of other abundant nitriles such as HCN, HNC, and HC$_3$N.}
{}
{}
{}
{}


\maketitle

\section{Introduction}

Nitriles, i.e., molecules containing a functional group $-$C$\equiv$N, are present in diverse astronomical environments. In particular, cyanopolyynes, H$-$(C$\equiv$C)$_n-$C$\equiv$N, are commonly found in cold interstellar clouds (\cite{bro1978} 1978; \cite{bel1997} 1997), circumstellar envelopes around evolved stars, especially in carbon-rich objects (\cite{win1978} 1978; \cite{par2005} 2005), and planetary atmospheres with a high content of carbon and nitrogen, as that of Titan (\cite{kun1981} 1981; \cite{cou1991} 1991).

It has been suggested that dicyanopolyynes, molecules containing two cyano groups, N$\equiv$C$-$(C$\equiv$C)$_n-$C$\equiv$N, could be abundant in interstellar and circumstellar clouds (\cite{kol2000} 2000; \cite{pet2003} 2003). However, these molecules cannot be detected through their rotational spectrum because they lack a permanent electric dipole moment. The simplest member of this series, cyanogen (NCCN), is thought to be a major precursor of the CN radical observed in cometary comae (e.g., \cite{fra2005} 2005), although its detection in a comet still remains challenging. Moreover, infrared observations carried out with the Voyager 1 have identified NCCN in the atmosphere of Titan (\cite{kun1981} 1981) and the larger homologue NC$_4$N has long been thought to be present as well (\cite{jol2015} 2015). It is also worth noting that a chemical cousin of cyanogen in which one N atom is substituted by a P atom, NCCP, has been tentatively identified in the C-rich envelope IRC\,+10216 (\cite{agu2014} 2014).

Given the difficulty to directly detect NCCN and larger dicyanopolyynes in cold interstellar clouds, it has been proposed that indirect strategies to probe their presence would be to observe chemically related molecules such as the polar metastable isomer CNCN or the protonated form NCCNH$^+$ (\cite{pet2003} 2003). Here we report the first detection in the interstellar medium of NCCNH$^+$, the protonated form of cyanogen, toward the cold dark clouds TMC-1 and L483.

\section{Observations}

Protonated cyanogen is a highly polar linear molecular cation. Its rotational spectrum has been characterized in the laboratory from microwave to millimeter wavelengths (\cite{ama1991} 1991; \cite{got2000} 2000) and its electric dipole moment has been calculated as 6.448~Debye (\cite{bot1990} 1990). We have observed the $J=5-4$ transition at 44.4~GHz with the Yebes 40m telescope and the $J=10-9$ transition at 88.8~GHz with the IRAM 30m telescope toward the cold dark clouds TMC-1 and L483 at the positions observed by \cite{mar2005} (2005, 2007) and \cite{agu2008} (2008), respectively.

\subsection{IRAM 30m}

The IRAM 30m observations were carried out using the EMIR 3 mm receiver and the fast Fourier Transform spectrometer with a spectral resolution of 50~kHz. The frequency switching technique was used to optimize the telescope time. The half power beam width (HPBW) of the IRAM 30m telescope at 88.8~GHz is 27\secp3. The observations of TMC-1 are part of a spectral line survey at 3 mm (\cite{mar2005} 2005, 2007, 2009). Most of the observations at 88.8 GHz used in this article were taken in February 2012, when the good weather conditions resulted in a system temperature of $\sim70$~K. More details are given in \cite{cer2012} (2012). The observations of L483 were taken from September to November 2014 during an observational campaign aimed at observing negative ions in dense molecular clouds and are described in detail in \cite{agu2015} (2015).

\begin{table}
\caption{Observed line parameters of NCCNH$^+$ in TMC-1 and L483} \label{table:lines}
\centering
\begin{tabular}{lcccc}
\hline \hline
\multicolumn{1}{c}{Transition} & \multicolumn{1}{c}{Frequency} & \multicolumn{1}{c}{$V_{\rm LSR}$} & \multicolumn{1}{c}{$\Delta v$}      & \multicolumn{1}{c}{$\int T_A^* dv$} \\
                                                & \multicolumn{1}{c}{(MHz)}                 & \multicolumn{1}{c}{(km s$^{-1}$)}    & \multicolumn{1}{c}{(km s$^{-1}$)} & \multicolumn{1}{c}{(K km s$^{-1}$)} \\
\hline
\multicolumn{5}{c}{TMC-1} \\
\hline
$J=5-4$  & 44379.850 & +5.68(8) & 0.72(13) & 0.011(2) \\ 
$J=10-9$ & 88758.108 & +5.91(9) & 0.48(12) & 0.006(1) \\ 
\hline
\multicolumn{5}{c}{L483} \\
\hline
$J=5-4$  & 44379.850 & +5.11(5) & 0.26(4)  & 0.006(1) \\ 
$J=10-9$ & 88758.108 & +5.31(9) & 0.49(13) & 0.017(4) \\ 
\hline
\end{tabular}
\tablenotea{\\
Numbers in parentheses are 1$\sigma$ uncertainties in units of the last digits.
}
\end{table}

\begin{figure}
\centering
\includegraphics[angle=0,width=\columnwidth]{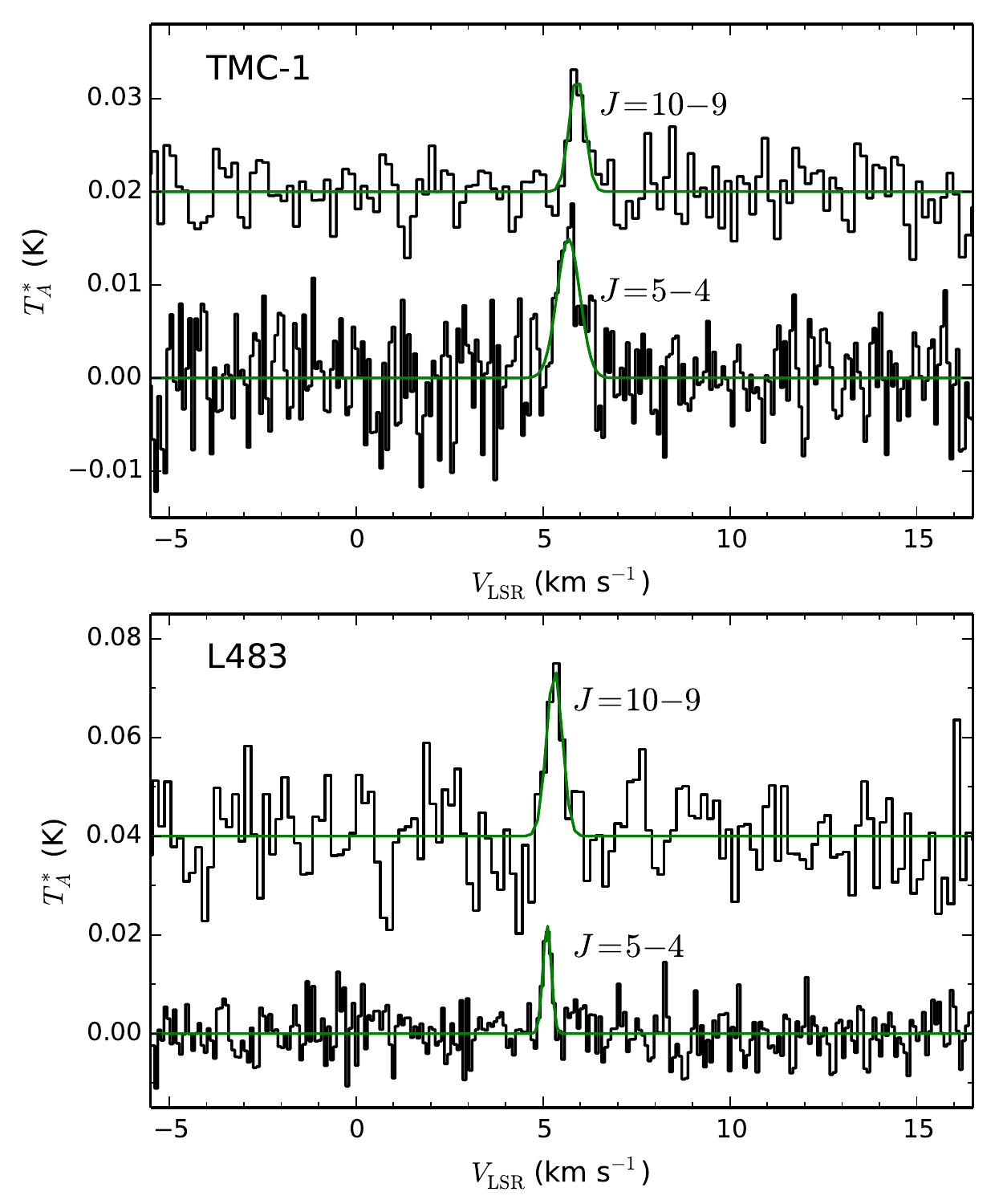}
\caption{Spectra of TMC-1 and L483 showing the emission lines assigned to the $J=5-4$ and $J=10-9$ rotational transitions of NCCNH$^+$, lying at 44.4~GHz and 88.8~GHz, respectively. The $T_A^*$ rms of the spectra at 44.4~GHz is 0.0045~K per 12~kHz channel for both sources, while at 88.8~GHz the $T_A^*$ rms is 0.0029~K and 0.0088~K per 50~kHz channel, for TMC-1 and L483, respectively. The antenna temperature scale can be converted to main beam brightness temperature by dividing by ($B_{\rm eff}/F_{\rm eff}$), which takes a value of 0.36/0.90 for the Yebes 40m telescope at 44.4~GHz and of 0.82/0.90 for the IRAM 30m telescope at 88.8~GHz.} \label{fig:lines}
\end{figure}

\subsection{Yebes 40m}

The 40m radiotelescope is located at Yebes (Guadalajara, Spain) at 980~m above sea level. Weather conditions are typically dry; the ammount of precipitable water vapor ranges from 4~mm in winter to 14~mm in summer. The antenna has an homological design and its optics is that of a Nasmyth radiotelescope with several receivers in the Nasmyth cabin. The antenna is equipped with a cryogenic single pixel dual polarization 45~GHz receiver built at Yebes, whose instantaneous bandwidth ranges between 41 and 49~GHz. The IF only processes bands of 500~MHz and/or 1500~MHz and its signal is injected into a fast Fourier Transform spectrometer with several modules, which can be configured to provide 16384 channels on a bandwidth of 100~MHz or 500~MHz. Observations reported in this article used two modules 100~MHz wide with a spectral resolution of 6~kHz. Spectra were later on smoothed to a spectral resolution of 12~kHz.

The observations toward TMC-1 and L483 were carried out in 9 and 12 periods, respectively, of 5-7~h during March and April 2015. System temperatures ranged between 120 and 200~K. On-off observations were done with an integration time of 60 s and the off position located 600$''$ away in right ascension. Calibration was repeated every 20 minutes and consisted of observing the sky and a hot load at ambient temperature. The sky opacity was estimated using the weather conditions measured 400 m away from the antenna and the atmospheric transmission model ATM (\cite{cer1985} 1985; \cite{par2001} 2001). We believe that the sky opacity has a maximum error of 10~\% (estimated by comparing with skydip measurements). To keep a good pointing and focus, continuum observations were performed every 20 minutes towards UOri (for TMC-1) and OH26.5+0.6 (for L483). Pointing and focus are estimated to be accurate within $5''$. The HPBW of the Yebes 40m telescope at 44.4~GHz is 42\secp6.

The final spectra of TMC-1 and L483 were obtained by averaging 2440 and 2338 individual spectra, respectively, half of them with left(right) circular polarization. The intensity scale is given in antenna temperature ($T_A^*$), which corrects for the sky opacity and forward efficiency (90 \%). The main beam efficiency has been estimated to be 36~\% in an elevation range between 20$^\circ$ and 80$^\circ$ from observations toward Venus and Saturn. Calibration can be considered correct within a 10~\% of uncertainty, mostly coming from the aperture efficiency and atmospheric conditions.

\section{Results}

The emission lines observed toward TMC-1 and L483 and assigned to the $J=5-4$ and $J=10-9$ transitions of protonated cyanogen are shown in Fig.~\ref{fig:lines}. Line parameters derived from Gaussian fits using GILDAS are listed in Table~\ref{table:lines}. The observations indicate that in TMC-1 the $J=5-4$ line is more intense than the $J=10-9$, while in L483 the contrary is found. This fact suggests that NCCNH$^+$ has probably a higher rotational temperature and/or a more compact spatial distribution in L483 than in TMC-1. The $J=5-4$ and $J=10-9$ transitions have upper level energies of 6.4~K and 23.4~K, respectively, and are observed with main beam sizes of 42\secp6 and 27\secp3, respectively.

Since we do not have information on the spatial distribution of NCCNH$^+$ in these two sources, we may assume that the emission size fills the main beam of the IRAM 30m and Yebes 40m telescopes to derive a first order estimate of the column densities of protonated cyanogen. In TMC-1, we derive a beam-averaged column density of $(8.6 \pm 4.4) \times 10^{10}$~cm$^{-2}$ and a rotational temperature of $5.8 \pm 0.8$~K. In L483, we derive a somewhat smaller column density, $(3.9 \pm 1.8) \times 10^{10}$~cm$^{-2}$, and a higher rotational temperature, $13.4 \pm 4.5$~K, which however has an important degree of uncertainty. If we adopt H$_2$ column densities of $1\times10^{22}$~cm$^{-2}$ in TMC-1 (\cite{cer1987} 1987) and $3\times10^{22}$~cm$^{-2}$ in L483 (\cite{taf2000} 2000; see also \cite{agu2015} 2015), we end up with fairly low fractional abundances relative to H$_2$, $8.6\times10^{-12}$ in TMC-1 and $1.3\times10^{-12}$ in L483.

\section{Discussion}

The detection of NCCNH$^+$ is a good indication of the presence of NCCN in dark clouds. However, to have an idea of how abundant is cyanogen in this type of sources it is necessary to have a look at the chemistry of protonated molecules in general, and NCCNH$^+$ in particular. In a simplified chemical scheme, a protonated molecule MH$^+$ can be formed by proton transfer to M
\begin{equation}
\rm XH^+ + M \rightarrow MH^+ + X, \label{reac:proton-transfer}
\end{equation}
where M must have a higher proton affinity than X in order to make the reaction exothermic. On the other hand, in cold dark clouds molecular cations are usually depleted by dissociative recombination with electrons
\begin{equation}
\rm MH^+ + e^- \rightarrow products, \label{reac:dissociative-recombination}
\end{equation}
where the products can be assorted neutral fragments. Within this simpe chemical scheme, at steady state we have
\begin{equation}
{\rm \frac{[MH^+]}{[M]}} = \frac{k_{PT}}{k_{DR}} \rm{\frac{[XH^+]}{[e^-]}}, \label{eq:abundance-ratio}
\end{equation}
where $k_{PT}$ and $k_{DR}$ are the rate constants of the reactions of proton transfer and dissociative recombination, respectively. Eq.~(\ref{eq:abundance-ratio}) suggests that high proton affinities of M would tend to enhance the rate of proton transfer (by increasing the number of available proton donors XH$^+$ and probably the associated rate constants $k_{PT}$) and so the abundance ratio [MH$^+$]/[M]. The above chemical scheme may however be not suitable for all protonated molecules, making necessary to build a chemical model that includes all relevant reactions to provide more precise predictions.

\begin{figure}
\centering
\includegraphics[angle=0,width=\columnwidth]{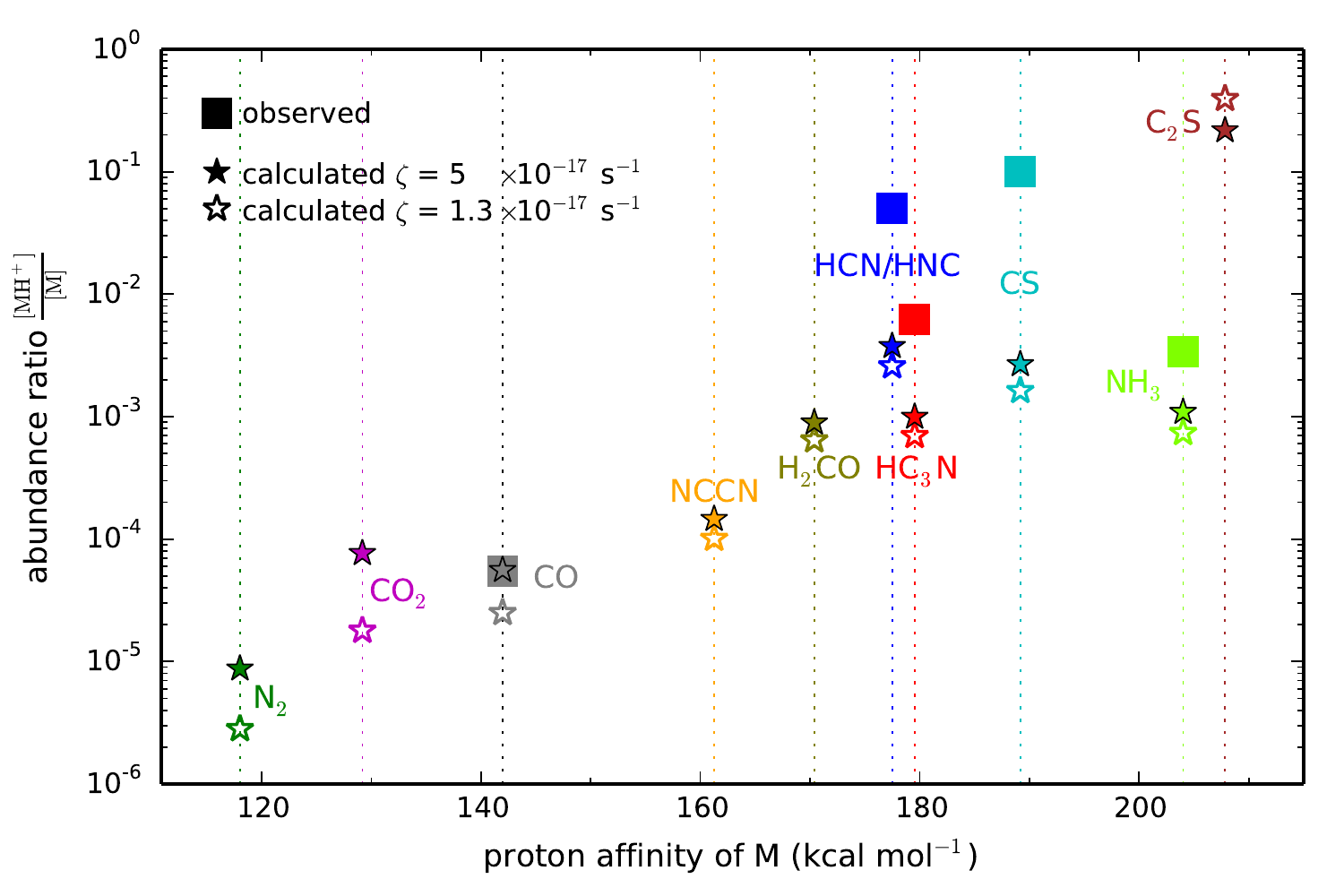}
\caption{Abundance ratios [MH$^+$]/[M] between the protonated and non protonated form of some molecules as a function of the proton affinity of M. Calculated values at steady state (reached after some 10$^5$ yr; see Fig.~\ref{fig:abun}), adopting two different cosmic-ray ionization rates are compared with values derived from observations of TMC-1 (\cite{agu2013} 2013). In the case of NH$_4^+$, the observed value is derived from observations of the monodeuterated species in Barnard 1 (\cite{cer2013} 2013). The values for HCN/HNC include both isomers and are plotted at the mean value of the proton affinities of HCN and HNC, 170.4~kcal mol$^{-1}$ and 184.6~kcal mol$^{-1}$, respectively. Note the trend of increasing [MH$^+$]/[M] with increasing proton affinity of M.} \label{fig:ratios}
\end{figure}

We have built a pseudo time-dependent gas phase chemical model of a dark cloud adopting standard physical parameters ($T_k=10$~K, $n_{\rm H}=2\times10^4$~cm$^{-3}$, $\zeta=1.3\times10^{-17}$~s$^{-1}$, $A_V=30$) and ``low metal'' elemental abundances (see \cite{agu2013} 2013). We have adopted the UMIST {\small RATE12} reaction network (\cite{mce2013} 2013) with a subset of reactions involving HCCN from \cite{loi2015} (2015). We find that the chemical model validates the leading role of dissociative recombination, reaction~(\ref{reac:dissociative-recombination}), as the major destruction process of protonated molecules, except for N$_2$H$^+$ and HCO$_2^+$, which are also depleted to an important extent by proton transfer to CO. We also find that, depending on whether the proton affinity of M is below or above that of CO, the main proton donor XH$^+$ in reaction~(\ref{reac:proton-transfer}) is either H$_3^+$ or HCO$^+$, respectively. However, some protonated molecules such as HCO$_2^+$, HCS$^+$, and HC$_2$S$^+$ are mainly formed by ion-neutral processes other than reaction~(\ref{reac:proton-transfer}), in which case Eq.~(\ref{eq:abundance-ratio}) underestimates the abundance ratio [MH$^+$]/[M].

Appart from NCCNH$^+$ and the widespread ions HCO$^+$ and N$_2$H$^+$, some other few protonated molecules have been observed in cold dark clouds: HCS$^+$ (\cite{tha1981} 1981), HCNH$^+$ (\cite{sch1991} 1991), HC$_3$NH$^+$ (\cite{kaw1994} 1994), HCO$_2^+$ (\cite{tur1999} 1999; \cite{sak2008} 2008), and NH$_3$D$^+$ (\cite{cer2013} 2013). The abundance ratio [MH$^+$]/[M], when available from observations, is well reproduced by the chemical model for M = CO and NH$_3$, although it is underestimated for M = HCN/HNC, HC$_3$N, and especially CS (see Fig.~\ref{fig:ratios}). Taking into account that HCNH$^+$, HC$_3$NH$^+$, and HCS$^+$ are mainly destroyed by dissociative recombination with electrons, whose rate constants are well known from experiments (\cite{sem2001} 2001; \cite{gep2004} 2004; \cite{mon2005} 2005), and that the rate constants of formation by proton transfer from HCO$^+$ are usually well constrained experimentally (\cite{ani2003} 2003), it is likely that the underestimation occurs because the chemical model misses important formation routes to HCNH$^+$, HC$_3$NH$^+$, and HCS$^+$.

\begin{figure}
\centering
\includegraphics[angle=0,width=\columnwidth]{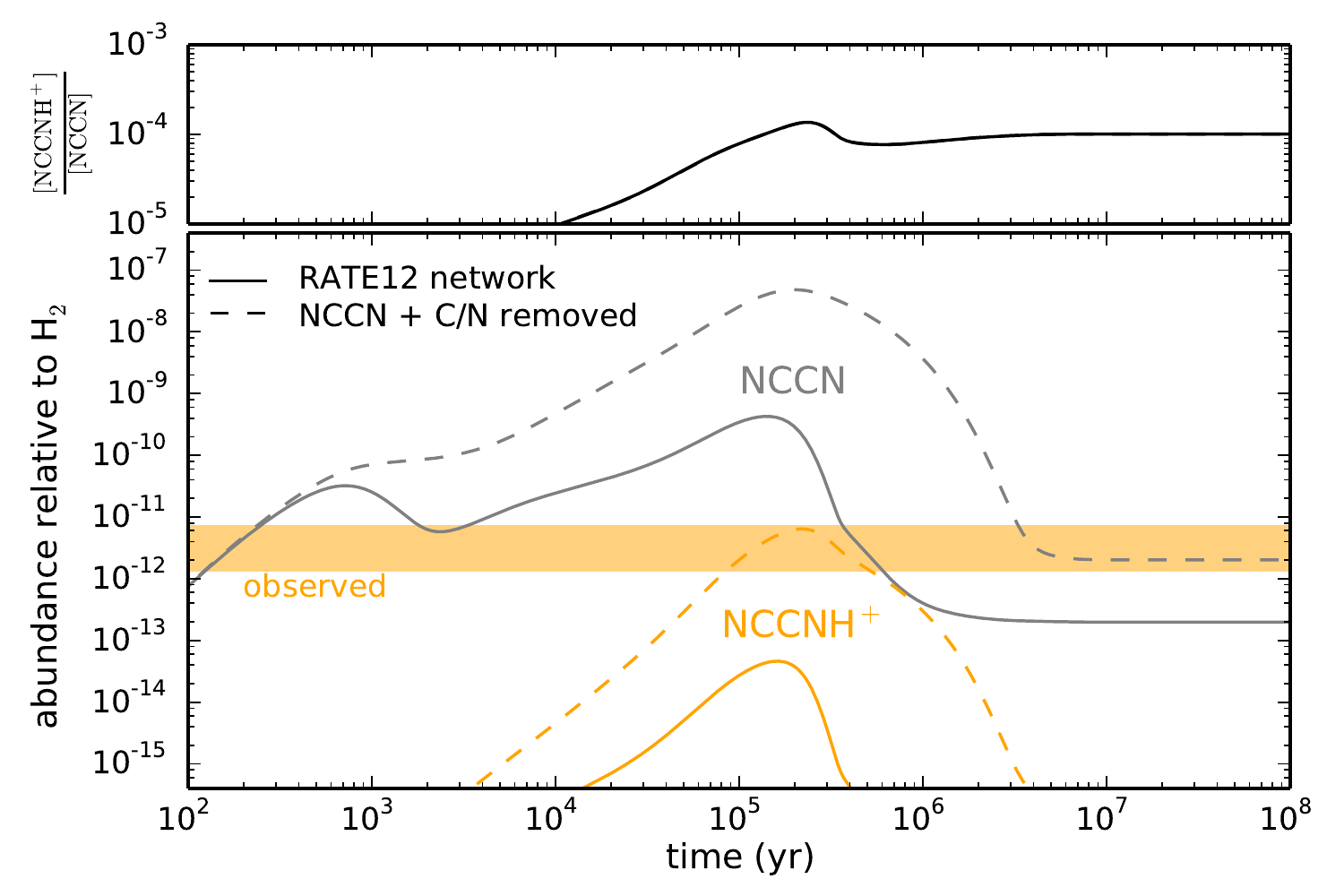}
\caption{Calculated abundances of NCCNH$^+$ and NCCN as a function of time using the UMIST {\small RATE12} chemical network (solid lines) and after removing the reactions of destruction of NCCN with C and N atoms (dashed lines), in both cases adopting $\zeta=1.3\times10^{-17}$ s$^{-1}$. The range of abundances of NCCNH$^+$ observed in TMC-1 and L483 is also indicated. The calculated abundance ratio [NCCNH$^+$]/[NCCN] is shown in the upper panel as a function of time.} \label{fig:abun}
\end{figure}

The abundance ratios [MH$^+$]/[M] are sensitive to the degree of ionization, and thus, to various physical parameters such as the cosmic-ray ionization rate $\zeta$ (the sensitivity is more marked for low proton affinities; see Fig.~\ref{fig:ratios}) and the volume density of particles (the higher the density, the lower the ionization fraction and thus the lower the importance of protonated molecules). It is interesting to note that both the chemical model and the observations suggest a trend in which the abundance ratio [MH$^+$]/[M] increases with increasing proton affinity of M. Since the destruction of MH$^+$ is controlled by dissociative recombination with electrons, whose rate constants are usually similar within one order of magnitude, the trend can be explained in terms of an enhanced formation rate of MH$^+$ with increasing proton affinity of M, resulting from the appearance of multiple formation pathways to MH$^+$ through exothermic ion-neutral reactions of proton transfer or other type. We may therefore expect to find other abundant protonated molecules MH$^+$ in dark clouds, as long as M has a high proton affinity and a sufficiently high abundance. Molecules such as H$_2$COH$^+$ (previously detected by \cite{ohi1996} (1996) toward warm, but not cold, clouds), OCSH$^+$, and C$_2$H$_3^+$ are potentially detectable. However, the most promising candidate is probably HC$_2$S$^+$, given that it has a high proton affinity and dipole moment (\cite{bar1991} 1991; \cite{mac1992} 1992; \cite{puz2008} 2008) and it is predicted to be just a few times less abundant than C$_2$S, whose column density in dark clouds is in the range (1-$100)\times10^{12}$ cm$^{-2}$ (\cite{fue1990} 1990; \cite{suz1992} 1992).

The chemical model predicts a low abundance of NCCNH$^+$ (see Fig.~\ref{fig:abun}), although it is not clear whether it is because the model underestimates the abundance of NCCN or the abundance ratio [NCCNH$^+$]/[NCCN]. According to the chemical model, NCCNH$^+$ is formed by proton transfer from HCO$^+$ to NCCN and destroyed by dissociative recombination with electrons, the rate constants of which are not known, although they are unlikely to be radically different from the values guessed in the UMIST {\small RATE12} reaction network. We thus expect the abundance ratio [NCCNH$^+$]/[NCCN] to be around 10$^{-4}$, as predicted by the chemical model, or higher, if important formation routes to NCCNH$^+$ are missing in the model. If the abundance ratio [NCCNH$^+$]/[NCCN] is correctly predicted, then the low fractional abundance predicted for NCCNH$^+$ must arise from an underestimation of the abundance of NCCN. In the chemical model, cyanogen is essentially formed by the reaction 
\begin{equation}
\rm HNC + CN \rightarrow NCCN + H, \label{reac:hnc+cn}
\end{equation}
for which \cite{pet2003} (2003) estimate a rate constant of $2\times10^{-10}$~cm$^3$~s$^{-1}$, and depleted by reaction with C and N atoms, whose rate constants at room temperature are inferred to be a few 10$^{-11}$~cm$^3$~s$^{-1}$ (\cite{why1983} 1983; \cite{saf1968} 1968). The calculated abundance of NCCN (and NCCNH$^+$) is very sensitive to these rate constants (see Fig.~\ref{fig:abun}). Therefore, a better understanding of the low temperature chemical kinetics of reaction~(\ref{reac:hnc+cn}) and the reactions of NCCN with C and N atoms seems the most immediate step to better constrain the chemistry of cyanogen in cold interstellar clouds. Additional uncertainties could come from other reactions that affect the precursors HNC and CN. For example, the systematic use of enhanced rate constants for most ion-polar reactions, adopted in the UMIST {\small RATE12} network, results in lower abundances for HNC, NCCN, and NCCNH$^+$, as compared with the use of a chemical network in which these rate constant enhancements are not adopted (see, e.g., \cite{woo2007} 2007). Formation of cyanogen on the surface of dust grains followed by some non-thermal desorption process is also a possibility to explore.

If we trust the abundance ratio [NCCNH$^+$]/[NCCN] of $\sim10^{-4}$ calculated by the chemical model, the abundance of NCCN in dark clouds could be as high as (1-$10)\times10^{-8}$ relative to H$_2$, i.e., comparable to that of other abundant nitriles such as HCN, HNC, and HC$_3$N.

\acknowledgements

We thank the IRAM 30m staff for their help during the observations and our referee, John Black, for useful comments. M.A. and J.C. acknowledge funding support from the European Research Council (ERC Grant 610256: NANOCOSMOS) and from Spanish MINECO through grants CSD2009-00038, AYA2009-07304, and AYA2012-32032. Authors from Yebes Observatory acknowledge funding by MINECO through grant FIS2012-32096. E.R. and M.G. acknowledge funding support from the CNRS program "Physique et Chimie du Milieu Interstellaire" (PCMI).

\end{document}